\documentclass[reprint, amsmath,amssymb,
 aps]{revtex4-2}

\usepackage{graphicx}% Include figure files
\usepackage{dcolumn}% Align table columns on decimal point
\usepackage{bm}% bold math
\usepackage{float}
\usepackage{multirow}
\usepackage{comment}
\usepackage{lipsum}
\begin{document}

%\preprint{APS/123-QED}

%\title{Multipole Hydrodynamic Interactions in Two-Dimensional Active Systems}
\title{Hydrodynamically Induced Aggregation in Two-Dimensional Active Systems}
\author{Roee Bashan and Naomi Oppenheimer}
 \affiliation{School of Physics and Astronomy and the Center for Physics and Chemistry of Living Systems, Tel Aviv University}
\date{\today}% It is always \today, today,

\begin{abstract}
%Ensembles of active particles often display complex collective dynamics. 
We investigate a system of co-oriented active particles interacting only via hydrodynamic and steric interactions. We offer a new method of calculating the flow created by any active particle in a 2D fluid, focusing on the dynamics of flow fields with a high-order spatial decay, which we analyze using a geometric Hamiltonian. We show that when orientational degrees of freedom are quenched, and the flow has a single, odd power decay, such many-particle systems lead to stable, fractal-like aggregation, with the only exceptions being the force dipole. We discuss how our results can easily be generalized to more complicated force distributions and to other effective two-dimensional systems.
\end{abstract}
\maketitle

Out-of-equilibrium ensembles, both biological and artificial, are often suspended in a fluid. Many times, these active systems have lower dimensions \cite{lauga2006swimming, lushi2014bacteria, spagnolie2019active2D,  bagaria2022dynamics}. A prominent example is a cell membrane --- an effective two-dimensional (2D) fluid in which both passive and active proteins are immersed \cite{membrane, oppenheimer2009correlated}. Active membrane proteins act as either rotary proteins, such as ATP synthase \cite{lenz2003rotlet, oppenheimer2019Crystalization, oppenheimer2022hyperuniformity}, or as shakers  --- particles that are not self-propelled but apply active forces on the membrane due to conformational changes, polymerization, or reorganization \cite{manikantan2020collective}. Artificial 2D systems are also widespread, such as active colloids driven by a chemical reaction~\cite{swimmers-colloids1, swimmers-light1, swimmers-light2}, by light~\cite{ben2022cooperation, modin2023hydrodynamic} or by an external magnetic field~\cite{swimmers-misc1}. Particle dynamics can be dominated by complex interactions (e.g. by electrostatics, capillary forces or an external field)~\cite{interaction1,interaction2}, or by purely hydrodynamic flows \cite{hydro-interactions1, big_review, saintillan2013active, rallabandi2019motion}.
In all the above systems, understanding the connection between flow and structure is crucial. 

The flow created by an active particle can be described by a multipole expansion (similar to electrostatics). Far away from the particle, the leading order of the flow, the stokeslet, is given by the total force the particle applies. Closer to the particle, higher multipole contributions start to dominate. First, the force dipole, which could be divided to the rotlet ~\cite{lenz2003rotlet, Lushi2015, yeo2015collective, samanta2021vortex, oppenheimer2022hyperuniformity}, and the stresslet ~\cite{deGraaf2017microswimmer,shoham2023swimmers,kos2018elementary, lauga2021zigzag}. Closer still, the quadrupolar term is significant, then the octopole, and so forth \cite{kim2013microhydrodynamics, multipole_example, spagnolie2012hydrodynamics}. Thus, by first finding the effect of a single point-force, the so-called Green's function of the problem, we can derive the general flow response. This procedure applies even when there is no net force, in which case the higher orders dominate at larger distances.

\begin{figure*}[tbh]
    \centering
    \includegraphics[width=0.9\linewidth]{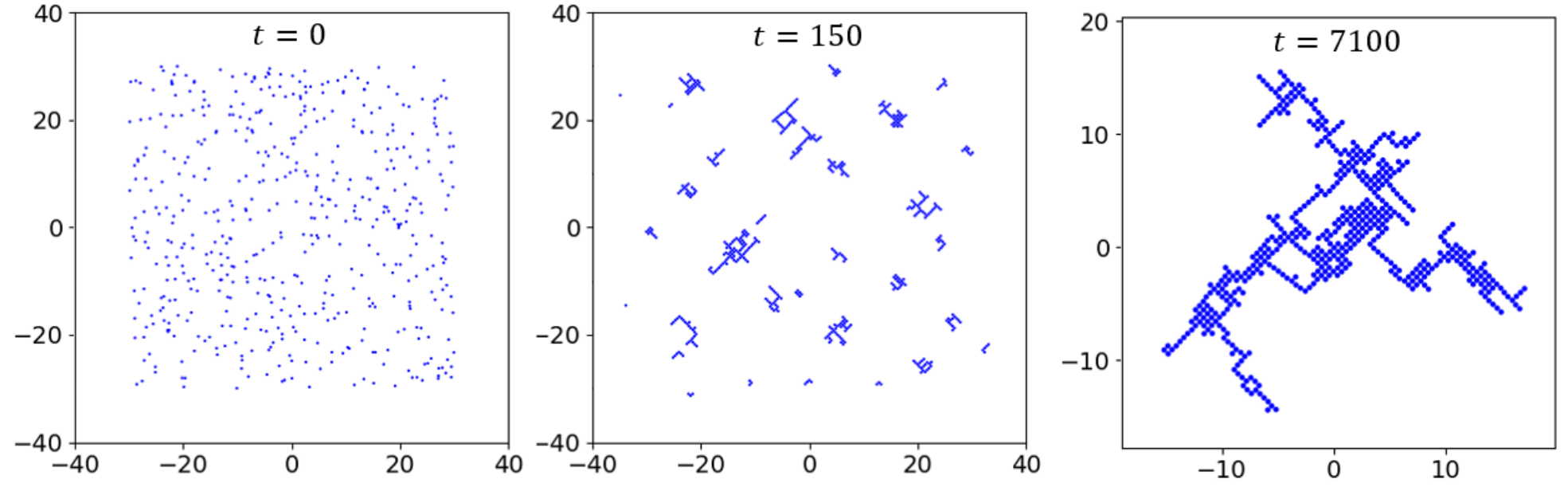}
    \caption{Snapshots of a simulation of 500 active particles with an octopole force distribution (i.e., the velocity decays as $1/r^3$). Particles are initialized randomly in a $60\times60$ box; Their orientation is quenched. The particles collide quickly only due to hydrodynamic interactions and form small clusters. At long times, clusters also collide, creating larger clusters. Finally, The particles form a single, fractal-like, cluster organized in a $45^\circ$ angled square lattice.}
    \label{fig:sim}
\end{figure*}

In this work, we explore the dynamics of suspended particles in a 2D viscous fluid with quenched orientational degrees of freedom, interacting only via the flow they create at their boundary and steric interactions. We review how the multipole terms mentioned above are often calculated from the stokeslet, but then, we focus on improving the common procedure, making higher-order terms easier to use and calculate. We show that, in many cases, namely for all even terms of the multiple expansion, the dynamics could be described by a Hamiltonian. This Hamiltonian is geometric in nature, with the conjugate variables being $x_i$ and $y_i$, the position of the $i^{\rm th}$ particle. Phase space in such cases corresponds to configurational space. Thus, by limiting possible paths in phase-space we can determine the steady-state distributions such active particles can take. We prove that, in many cases, particles aggregate. In particular, in all far-field limits, which are dominated by a single multipole term with an odd power decay. The only exceptions are the rotlet and the stresslet. We corroborate our findings by simulating particles that interact by an octopolar force distribution. Figure~(\ref{fig:sim}) illustrates our key finding: a pure multipole force distribution leads to particle aggregation. Moreover, we claim that our findings can be applied to more general 2D cases --- both for calculating high-order interactions in more complex fluids, and for understanding their resulting dynamics in many-particle ensembles.

The Reynolds number of a flow, $\text{Re}=\frac{\rho UL}{\mu}$, is the ratio between the inertia of the fluid and the viscous forces in its flow, where $\rho$ is the fluid's density, $\mu$ is its viscosity, $U$ is the characteristic velocity and $L$ the characteristic length. For microswimmers in biological systems, the Reynolds number is small, as the characteristic lengths are microscopic. In the low Reynolds number limit, the inertia is thus negligible, and the flow velocity $\textbf{v({\bf r})}$ (where ${\bf r} = (x,y)$ is the 2D position vector) is governed by the Stokes equations, 
\begin{equation}
\begin{cases}
    \nabla p = \mu \nabla ^2 \textbf{v} + \textbf{f}\\
    \nabla \cdot \textbf{v} =0,
\end{cases}
\label{eq:stokes}
\end{equation}
where $\textbf{f}$ is an external force. From now on, we will work in unitless notation where $\mu=1$. The flow due to a point force, the stokeslet, also termed the Oseen tensor in 2D \cite{oseen}, is given by
\begin{equation}
\label{eq:oseen}
    G^{\rm 2D}_{ij}=\frac{1}{4\pi}\left(-\log\frac{r}{r_{0}}\delta_{ij}+\frac{r_ir_j}{r^{2}}\right),
\end{equation}
where $r= |\textbf{r}|$, and $r_0$ is a normalizing constant, e.g. the size of the system (it is insignificant for our purposes since it does not appear in higher order terms). This solution is the Green's function of the Stokes equations and can be used to calculate the flow due to any force distribution. However, an exact calculation is generally difficult, and, when particles are far apart, unnecessary. Instead, one often approximates the flow in powers of $r$ using a Taylor expansion of $G^{\rm 2D}$, giving a solution of the form, 
\begin{equation}
\label{eq:monopoles approximation}
    \textbf{v} = \bm{F}\cdot \bm{G} + \bm{D}_2 \boldsymbol{:}\nabla \bm{G} +\cdots+\bm{D}_N \raisebox{.5pt}{\textcircled{\raisebox{-.9pt} {N}}}\nabla^{\otimes N-1}\bm{G} + \cdots.
\end{equation}
where $\boldsymbol{:}$ is the second order inner product, \raisebox{.5pt}{\textcircled{\raisebox{-.9pt} {N}}} is the $N^{th}$ order inner product, $\otimes$ is the tensor product (which, when found in the exponent, refers to the times to apply the operator), $\textbf{F}$ is the total force, and $\textbf{D}_n$ is the $n^{\rm th}$ force coefficient given by 
\begin{equation}
    \bm{D}_{n+1} = \frac{(-1)^n}{n!}\int_S \bm{f}\otimes\bm{r}^{\otimes n} dA.
\end{equation}
%\begin{equation}
%    D_{ji_1i_2\dots i_n} = \int_S f_jr_{i_1}\dots r_{i_n} dA,
%\end{equation}
The integration is over the area where the forces are applied.
Using the monopole or dipole terms for solutions far away from the particle is often enough. Calculations of higher-order terms are messy at best. To evaluate the $N^{\rm th}$ term, first one needs to calculate the mixed $N-1$ order derivatives, then calculate $N$ coefficients of $\textbf{D}_N$, and finally contract these two organized in $N-1$ rank and $N$ rank tensors in a slaving procedure. The contraction result is often complicated in Cartesian coordinates but is much simpler in polar coordinates. A similar formulation is also used in 3D~  \cite{multipole_example}, with the equivalent $G^{\rm 3D}$. In Sec.~\ref{sec_oneParticle}, we will show a different route to calculating the flow using the biharmonic equation.

Let us now consider a fluid in which an active particle is immersed. The particle is propagated by the flow only (we will later add steric interactions), and so its position is given by a simple equation of motion $\dot{\textbf{r}}=\textbf{v}(\textbf{r})$. Since the flow is incompressible and has no sources of mass (Eq.~\ref{eq:stokes}), it can be described by a potential $\psi$ called the stream function. In 2D, $\nabla\times(\psi\hat{z})=\textbf{v}\equiv\nabla ^ \perp \psi$, where $\nabla ^ \perp = -\hat{z}\times\nabla = \hat{x}\partial_y-\hat{y}\partial_x$.
In the case of a flock of similar particles, all with the same orientation, each with strength $S_i$, the stream function created by all particles is a simple superposition, $\psi(\textbf{r}) =  \sum_i S_i\psi(\textbf{r}-\textbf{r}_i)$. The dynamics of the particles could be described by a geometric Hamiltonian 
\begin{equation}
\label{eq:hemiltonian}
    H=\frac{1}{2} \sum\limits_{i \neq j}S_iS_j\psi(\textbf{r}_i-\textbf{r}_j).
\end{equation}
The equations of motions are given by Hamilton equations of $H$, with $\sqrt{S_i}x_i$ and $\sqrt{S_i}y_i$ being conjugate variables. Such a Hamiltonian description has been useful for point vortices in an ideal fluid~\cite{hamilonian2, ideal_fluid} and also, more recently, in viscous systems~\cite{hamiltonian1,hamiltonian3,shoham2023swimmers,hamiltonian4, disk, bolitho2022overdamped}. However, this formalism only holds when $\psi$ is an even function of $\textbf{r}$. An odd component of $\psi$ will cancel out in the (symmetric) summation, and $H$ would be identically zero. In the rest of this work, we will mostly use the case where the particles are identical, i.e. $S_i=1$.

The rest of the paper is organized as follows. In Sec.~\ref{sec_oneParticle} we will derive the 2D flow created by an active particle with any force distribution acting on its surface. In Sec.~\ref{sec_twoParticles}, we will derive the dynamics of two similar particles with quenched orientational degrees of freedom. As an example, we will show that when the force distribution is octopolar, dominated by ${\bf D_4}$, particles will collide at a finite time. We will then prove that, except for a force-dipole, all interactions of a single, even, multipole term lead to aggregation in the two-particle system. We will also discuss more general interactions. Our arguments will also be applicable to other similar systems with flow fields not described by the equations of flow above. In Sec.~\ref{sec_many}, we will claim that the two-particle case also applies to an ensemble of many particles and show a simulation resulting in aggregation.

\section{flow due to a single particle}
\label{sec_oneParticle}
Here, we will derive the stream function generated by a general active particle in a 2D viscous fluid. For simplicity, we will assume a circular particle with a radius $R$, applying a force on the fluid at its boundary. However, the choice that forces are only applied on a circle does not limit the result of the stream function, and all the solutions in Eq. (\ref{eq:monopoles approximation}) will appear here as well (see Supplementary Information for more details). 
Taking the curl of the first equation in Eqs.~(\ref{eq:stokes}), and writing the velocity in terms of $\psi$, we arrive at an equation for the stream function,
\begin{equation}
\label{eq:main_DE}
   \nabla^2 \nabla^2 \psi = -\nabla^\perp \cdot \textbf{f} = \hat{z} \cdot (\nabla \times \textbf{f}).
\end{equation}
The force is only present at the surface of the particle. We will assume all the forces are parallel, giving a force of the form $\textbf{f}=\frac{1}{R}\delta(r-R)F(\theta)\textbf{f}_0$ where $\textbf{f}_0$ is a constant and $F(\theta)$ is an angular density distribution.
The choice that forces are parallel creates some restrictions, but they are easily resolved for a force distribution with multiple orientations (see SI). The homogeneous biharmonic equation has been used to study the flow created by active swimmers with varied boundary conditions \cite{crowdy2010two, crowdy2013stokes}. Here, we focus on the coupling between the force applied by a swimmer and the flow, which is given by the non-homogeneous biharmonic equation~(\ref{eq:main_DE}).

In order to get a solution equivalent to Eq.~(\ref{eq:monopoles approximation}), we match the boundary conditions on the flow at infinity and at $r=0$ (despite the particle having a small radius $R$), similar to the derivation of the Oseen Tensor \cite{oseen}. It is possible to solve Eq. (\ref{eq:main_DE}) by decomposing the force and the stream function in a Fourier series as they are periodic in $\theta$,
\begin{equation}
\label{eqFourierSeries}
    \psi(r,\theta) = \sum\limits_{n\in\mathbb{Z}} B_n(r)e^{in\theta} \ \ \ ,\ \ \  F(\theta) = \sum\limits_{n\in\mathbb{Z}} C_ne^{in\theta}.
\end{equation}
We will assume no force monopole, thus $C_0=0$. Using Eq.~(\ref{eqFourierSeries}) in Eq.~(\ref{eq:main_DE}), we get (see SI for more details),
\begin{equation}
\label{eq:semigreen}
    \hat{O}_n^2 B_n(r)=i \frac{\eta}{R} C_{n-1}\left[\frac{d}{dr}\delta(r-R)-\frac{n-1}{r}\delta(r-R)\right],
\end{equation}
where $\hat{O}_n=\frac{1}{r}\frac{d}{dr}\left(r \frac{d}{dr}\right)-\frac{n^2}{r^2}$ and $\eta$  is the force in complex notation $\eta = |\mathbf{f_0}|e^{-i\arg(\textbf{f}_0)}$. The solution is given only by the real part of Eq.~(\ref{eq:semigreen}). This equation implies that a Fourier component $n$ in the stream function $\psi$ is caused by a component $n-1$ of the force. For example, a force caused by two opposing normal point forces on the surface has only odd components, which leads to only even components in $\psi$. 
Every Fourier component, $B_n(r)$, in the stream function $\psi$ is a linear combination of four power laws in $r$: $r^n, r^{n+2}, \frac{1}{r^n}, \frac{1}{r^{n-2}}$, where only negative exponents will contribute at $r>R$ (see SI for a detailed solution). Outside the particle, at $r>R$ the solution is a linear combination of terms of the form $\ln r, \sin(2\theta), \sin(n\theta)/r^n, \sin(\theta)/r^{n-2}$ (ignoring phases), where the logarithm is due to a degeneracy for $n=0$. The complete solution outside the particles is given by
\begin{comment}
\label{eq:mega_solution}
\begin{aligned}
    & \frac{\psi}{R} = -\sum\limits_{n>1} i \eta C_{n-1} e^{in\theta} \left[ \frac{1}{4n}\left(\frac{R}{r}\right)^{n} +\frac{1}{4(1-n)}\left(\frac{r}{R}\right)^{2-n} \right] \\ 
    & - \sum\limits_{n<0} \frac{1}{4n(n-1)}i\eta C_{n-1}e^{in\theta}\left(\frac{r}{R}\right)^{n} +\frac{1}{2} i\eta C_{-1} \ln\frac{r}{R}.
\end{aligned}
\end{comment}
\begin{equation}
\label{eq:mega_solution}
\begin{aligned}
    & \frac{4\psi}{iR\eta}= \sum\limits_{n>1} \left\{ C_{n-1} e^{in\theta} \left[ \frac{1}{n}\left(\frac{R}{r}\right)^{n} +\frac{1}{(1-n)}\left(\frac{R}{r}\right)^{n-2} \right] \right. \\ 
    & +\left.\frac{C_{-n-1}}{n(n+1)}e^{-in\theta}\left(\frac{R}{r}\right)^{n}\right\} - 2 C_{-1} \ln\frac{r}{R} + \frac{1}{2} C_{-2} e^{-i\theta} \frac{R}{r}.
\end{aligned}
\end{equation}
This solution allows us to calculate the velocity field created by any active particle applying a force on the fluid in one direction. As mentioned earlier, this result is easily generalized to any force orientation as outlined in the SI. Furthermore, the terms in Eq.~\ref{eq:mega_solution} can apply to active particles that have different boundary conditions at the surface of the particle. For instance, a squirmer is an active particle defined by the flow on its surface, which has no radial component \cite{lighthill1952squirming, blake1971spherical}. Its 3D description is well-known. Similarly to Eq. (\ref{eq:mega_solution}), a general 2D squirmer is given by:
\begin{equation}
    \psi = B_0\ln\frac{r}{R}+\sum\limits_{n>1} B_n\left(\frac{R^{n}}{r^n}-\frac{R^{n-2}}{r^{n-2}}\right)\sin (n\theta +\phi_n)
\end{equation}
For any $B_n$ and $\phi_n$. These terms are of the same form as the different multiples of a 2D flow given in Eq.~\ref{eq:mega_solution}, i.e. $\sin(n\theta)/r^n, \sin(n\theta)/r^{n-2}$. %As an example, in Sec.~\ref{sec_octopole} we derive a pure octopolar flow.

\section{two particle dynamics}
\label{sec_twoParticles}
This section will focus on a particular example of the flow given by a single force multipole, as given by Eq.~ (\ref{eq:mega_solution}), and the dynamics of two particles interacting by such a flow. We will then discuss the dynamics of two particles interacting by a general solution to Eq.~(\ref{eq:main_DE}), with any desired force distribution. We will show that when the orientation is quenched, similar particles driven by a pure, even, multipolar term will always collide, the only exceptions being the force monopole and force dipole. For some general force distribution, two particles will always attract in the far-field limit, given that the total force is zero and there is no contribution from a force dipole. If there are no local minima and maxima in the stream function, the particles will also collide.

\subsection{Case Study: Octopole Term}
\label{sec_octopole}
\begin{figure}
    \centering
    \includegraphics[width=0.8\linewidth]{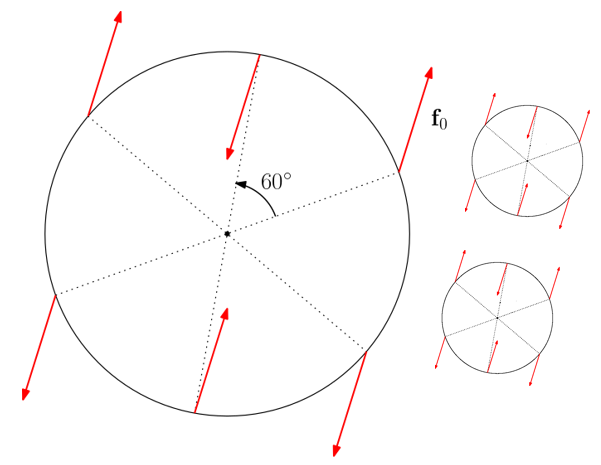}
    \caption{A realization of a particle whose stream function's leading term is $1/r^2$, which is the octopole term in Eq. (\ref{eq:monopoles approximation}), using point forces. We investigate the interaction between two identical particles with this force distribution, with a fixed orientation.}
    \label{fig:force}
\end{figure}

We will consider a force distribution such that only the third harmonic of the force distribution is present. In equation (\ref{eq:monopoles approximation}), this corresponds to the forth term, $\mathbf{D}_4$ dominating, and the stream function is given by:
\begin{equation}
    \psi_3= f_0C_3R \left[\left(\frac{R^2}{12r^2}-\frac{R^4}{16r^4}\right)\sin4\theta + \frac{R^2}{24r^2}\sin2\theta
    \right],
\end{equation}
assuming $C_3=C_{-3}$ is real, and the force is in the $x$ direction. In the far-field limit, taking $R\rightarrow0$ and $f_0\rightarrow\infty$ such that $R^3f_0$ remains constant, only the terms $~1/r^2$ remain, which are the octopole moment
\begin{equation}
\label{eq:psiOct}
    \psi_\text{oct}=\frac{R^3}{12r^2}C_3f_0 \left( \frac{1}{2}\sin2\theta+\sin4\theta\right).
\end{equation}
There are many force distributions that have this property. The obvious one is the pure harmonic written above. A more realistic realization given by point forces is presented in Fig. \ref{fig:force}. A distribution of forces on a disk can always be constructed so that any term is dominant in the multipole expansion (see SI).

\begin{figure}
    \centering
    \includegraphics[width=0.8\linewidth]{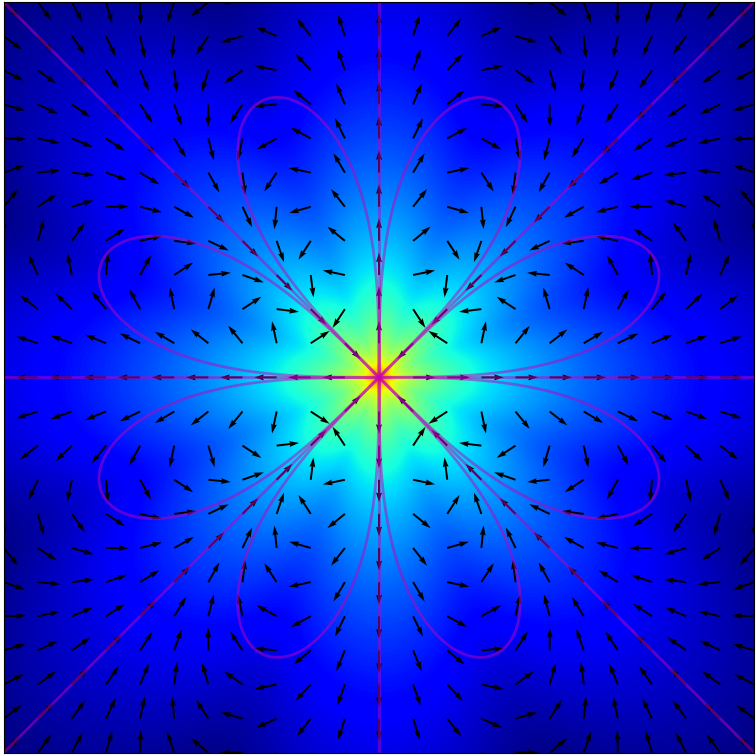}
    \caption{The vector field generated by a stream function $\psi = S\frac{\sin(4\theta)}{r^2}$, where blue corresponds to low strength and red is high. In purple we mark example paths that a particle can take. The straight paths are characterized by a zero Hamiltonian, and alternate being stable or unstable. The leaf-shaped paths are for the non-zero case, alternating a positive and negative Hamiltonian.}
    \label{fig:octopole}
\end{figure}

From this point on, for simplicity, we will focus on one of the terms in Eq.~\ref{eq:psiOct} and take $\psi = S\sin(4\theta)/r^2$ (This can be realized with a force distribution at two different directions, see SI), which represents well the general case for the dynamics caused by a single (even) term. Its corresponding flow is displayed in Fig.~\ref{fig:octopole}. A system of two identical particles creating such a flow is described by the Hamiltonian in Eq.~(\ref{eq:hemiltonian}). 

In Hamiltonian mechanics, symmetries of the Hamiltonian correspond to conservation laws. In this case, the Hamiltonian is symmetric to translation in time and space. Therefore, the Hamiltonian is conserved as well as, what we term, the ``center of activity", $(\textbf{r}_1+\textbf{r}_2)/2$, which is analogous to the center of mass. It is therefore enough to consider the dynamics of the relative difference between particles $\textbf{d}=\textbf{r}_1-\textbf{r}_2$, which has a 2D phase space. The paths $\textbf{d}$ can take are displayed in Fig. \ref{fig:octopole}, which are given by $r^2\propto\sin(4\theta)$ as a result of the conservation of the Hamiltonian $H=\psi(\textbf{d})$. It is also straightforward to solve the equations of motion directly, giving 
\begin{equation}
    \begin{cases}
    \cos(4\theta)=\cos(4\theta_0)-16\frac{H^2}{S}t\\
    r^4=r_0^4-32St\cos(4\theta_0)-256H^2t^2,
    \end{cases}
\end{equation}
where $r$ and $\theta$ are the polar coordinates in the phase space of $\textbf{d}$. Theta is constrained to only one of the intervals $(0,\frac{\pi}{4}),(\frac{\pi}{4},\frac{\pi}{2}) \cdots (\frac{7\pi}{4},2\pi)$ which are given by half the period of the harmonic. Also, in general, the radius does not decrease monotonically with time, however it always decreases to zero unless $H=0$. The special paths with $H=0$ are the only paths that will not necessarily converge to the origin. In principle, they can diverge to infinity, though a divergence is unstable as even a small perturbation to the Hamiltonian will cause the path to be bound.
Therefore, every stable solution of the pair leads to a collision. We will now prove that, in the far-field limit, this holds true in general, except for dipolar force terms.

\subsection{General dynamics of two Particles}
\label{sec_twoGeneral}
Let us consider a case of two identical particles, each creating a stream function $\psi$ that is described by a single (even) multipole term in the far-field limit. We will later consider a dipolar flow which is given by $\psi=\alpha\ln r+\beta \sin(2\theta+\phi)$. Any other flow is given by 
\begin{equation}
    \psi = \left[\alpha\sin(m\theta+\varphi)+\beta\sin((m+2)\theta+\phi)\right]\frac{R^m}{r^m}
\end{equation}
for some even $m$, any coefficients $\phi,\varphi$ and any non-trivial $\alpha$ and $\beta$. We will now show that all such flows (excluding the dipole terms) will result in a collision of the two particles, similar to the octopole term described above.

To prove this claim, we will examine the paths that the relative distance between particles, $\textbf{d}$, takes in its phase-space, whose motion is governed by $\psi$, $\dot{\textbf{d}}=2\nabla ^ \perp \psi(\textbf{d}) $. The ``grad perp" operator, $\nabla ^ \perp$, has an important geometrical interpretation: unlike the gradient, which points to the direction where a scalar function is most increasing, the ``grad perp" operator points to the direction perpendicular to that (anti-clockwise) which is also the direction where the function is constant. This interpretation is connected to the fact that $H$ is conserved along paths.

\begin{figure}
    \centering
    \includegraphics[width=0.8\linewidth]{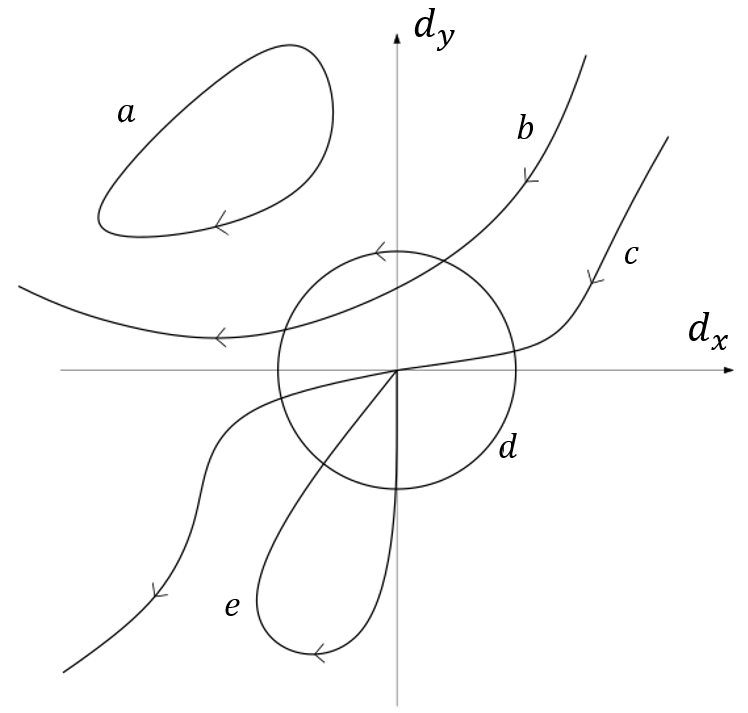}
    \caption{The categories of possible paths for the separation between two-active particles,  $\textbf{d}$. If bound, the path either encircles the origin, does not encircle it, or crosses it. If the path diverges, then it either crosses the origin or not. Each path is described by an equation $\psi (x,y) = H$. In the text we prove which paths are permitted and under what circumstances.}
    \label{fig:paths}
\end{figure}

First, we note that the stream function has no local maxima/minima, as a cross section of it at a constant $\theta$ gives $\psi\propto r^{-m}$, which never has local extrema. Thus, a point path is unstable. This is expected since a stable point would have a negative divergence of the velocity field. Thus, every path curve must be either closed or diverging to infinity. And, since $\psi(r \rightarrow \infty)=0$, every path diverging to infinity has a zero Hamiltonian, and every path with non-zero Hamiltonian is necessarily bound and closed. The different possible path categories are shown in Fig.~\ref{fig:paths}. We will now assess whether or not each is possible.

\textbf{Path \textit{a}}. A closed path that does not contain the origin, such as path {\it a}, encircles an area which is a compact subset of the phase-space, where $\psi$ is continuous. Because the stream function is constant along the edge of the region, that is, along a path, there must be a local minimum or maximum inside that region, contradicting our earlier conclusion (this is a higher dimensional case of Rolle's theorem~\cite{rol}). This argument, which disallows paths like {\it a}, does not hold for the closed paths like {\it d} and {\it e} because their closed area contains the origin, and so the stream function is not continuous on it.

\textbf{Paths \textit{b} and \textit{c}}.
Now consider a path that diverges to infinity, like paths {\it b} and {\it c}. As mentioned, such a path has zero Hamiltonian, so every path that crosses it must also have zero Hamiltonian. Therefore, the curve will split the plane in two --- paths on one side of the curve that have a non-zero Hamiltonian cannot cross to the other side. For paths like {\it b}, there is a side which does not contain the origin, and so closed paths contained in it must be like path {\it a}. Since we showed such paths are not allowed, it then follows that all paths in that side have zero hamiltonian, and thus $\psi=0$. We assume that there cannot be such an area in which the stream function is constant since that corresponds to an area with no flow. That means every path diverging to infinity must cross the origin, That is, it must be similar to path {\it c}. In fact, a path similar to path {\it c} must exist, because there is some $\theta_0\in[0,2\pi)$ such that $\psi(r,\theta_0)=0$ for all $r$, and so the ray $\theta=\theta_0$ is a path similar to path {\it c}.

\textbf{Path \textit{d}}. Since a divergent path such as {\it c} necessarily exists, every closed path encircling the origin (path {\it d}) crosses it. Thus, it also has a zero Hamiltonian. Here, again, the path splits the plane in two and forces paths similar to path {\it a}. Therefore, such a path as {\it d} cannot exist.

In conclusion, the only possible paths are either closed curves crossing the origin such as \textbf{path \textit{e}} or divergent curves with a zero Hamiltonian crossing the origin such as \textbf{path \textit{c}}. However, the side of the divergent path that goes to infinity as time increases is unstable. Consider a small perturbation to the Hamiltonian, $H=\varepsilon\neq0$, the particle's trajectory must now be bound. Thus, it is unstable and will turn to path {\it e}. And so, every stable path crosses the origin, which corresponds to the particles colliding.

This conclusion only applies to the far-field limit. As the particles get closer, eventually, higher-order terms of the flow will contribute to the dynamics. If the stream function $\psi$ has no minima or maxima, even in close-range, then by the same reasoning as above, the particles will collide because the only allowed paths are still {\it c} and {\it e}. However, if the stream function has an extremum in close range, then all path types are allowed. Particles will still aggregate up to these short distances. 

\begin{figure}
    \centering
    \includegraphics[width=0.72\linewidth]{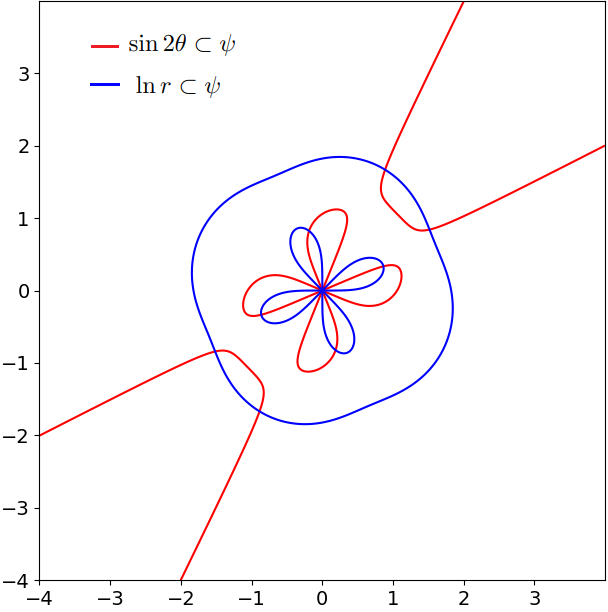}
    \caption{Phase space (configurational space) examples where dipole terms are allowed showing paths {\it b} and {\it d} exist in such cases. In blue, $\psi = 2\ln r + \frac{\sin 4\theta}{r^2}$ with $H=1.2$ and in red $\psi = \sin 2\theta + \frac{\cos 4 \theta}{r^2}$ with $H=0.8$. These paths are also stable, as a perturbation to the Hamiltonian gives a similarly shaped path.}
    \label{fig:dipole_paths}
\end{figure}

Let us now consider the different paths that the dipolar terms allow. If the stream function contains a dipolar term of the form $\sin2\theta $, then paths that diverge to infinity can have a non-zero Hamiltonian, and so paths like {\it b} are allowed. If the stream function contains the term $\ln r$, no paths can diverge to infinity, and so paths like {\it d} are allowed, while paths like {\it b} and {\it c} are not allowed. Examples for both are depicted in Fig. \ref{fig:dipole_paths}. Note that paths like path {\it a} are never allowed in the absence of extrema.

Lastly, we will address the dynamics when $\psi$ is not even. If the stream function is odd (the far-field limit always has a defined parity), then no Hamiltonian can be constructed. However, it is easy to see that now the relative distance, $\textbf{d}$, is conserved, while the ``center of activity" changes with velocity given by $\textbf{v}(\textbf{d})$. This results in the pair moving together at a constant velocity, and is true even in close range interaction. In general, if $\textbf{v} = \textbf{v}_O + \textbf{v}_E$ which are even and odd respectively, then
\begin{equation}
    \begin{cases}
    \dot{\textbf{r}}_{\text{cm}} = \textbf{v}_E(\textbf{d})\\
    \dot{\textbf{d}} = 2\textbf{v}_O(\textbf{d})
    \end{cases}
\end{equation}
and so the relative distance between particles, $\textbf{d}$, has the same dynamics as described earlier for an even stream function, as long as $\textbf{v}_O\neq0$ (note that $\textbf{v}$ is odd when $\psi$ is even and vice versa).

\section{many particle dynamics}
\label{sec_many}
\begin{figure}[tbh]
    \centering
    \includegraphics[width=0.8\linewidth]{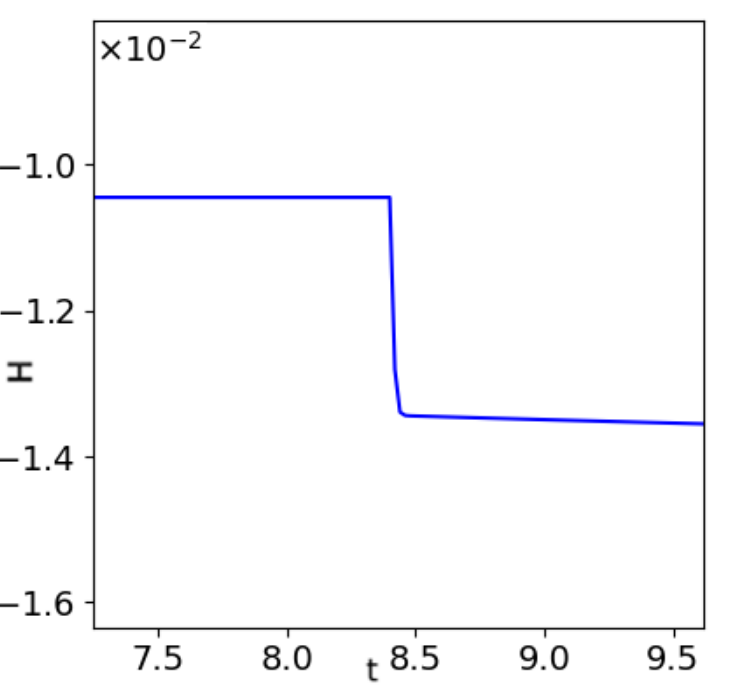}
    \caption{The Hamiltonian of a 4-particle system during a collision. A swift change in the Hamiltonian is observed at time $t = 8.5$ when a collision occurs.}
    \label{fig:collision}
\end{figure}

A system of many particles can no longer be solved analytically. However, for an even stream function it is possible to construct a Hamiltonian as in Eq.~(\ref{eq:hemiltonian}), from which follows the conservation of the center of activity $\sum_i \textbf{r}_i/N$ as well as $H$ itself. We integrate over the system's equations of motion using the python library \textit{scipy.integrate.DOP853}, which is an 8th order Runge-Kutta method. Since the particles are expected to collide, we also add steric interactions to the system, given by $\Delta \textbf{v}_{\text{steric}} = k_{\text{steric}} (l_{\text{steric}}-|\textbf{r}_{ij}|)\hat{\textbf{r}}_{ij}$ if $|\textbf{r}_{ij}|<l_{\text{steric}}$ and zero otherwise, where $\textbf{r}_{ij}$ is the distance from the i-th and j-th particles. The steric interaction gives a strong repelling force when particles overlap. In our simulation we used $l_{\text{steric}}=0.5$ and $k_{\text{steric}}=10^4$. The new interactions break the Hamiltonian description, and so $H$ is no longer  conserved. However, the center of activity should remain constant as the interaction is still symmetric. In addition, between collisions the Hamiltonian is still conserved (see Fig.~\ref{fig:collision}). The steric interactions are a highly simplified description of the close-range forces between active particles that are not described by stokes flow.

We initialize hundreds of randomly positioned particles. Each particle creates the velocity field discussed in Sec. III. Namely, we take an octopolar flow with the stream function $\psi = \sin4\theta/r^2$.
Since we have established the similarity of all far-field flows, excluding dipolar flow, as well as flows created by stream functions with no extrema, we can use this simulation to infer the behavior of ensembles interaction via such stream functions. The result of the simulation for different times is displayed in Fig. \ref{fig:sim}. 

Immediately pairs of particles start attracting and eventually collide. This behavior is a direct result of the two particle dynamics, because the stream function falls fast with $r$, $\psi \sim r^{-2}$, and so the majority of a particle's motion is determined by its closest neighbor, while the weak effect of other particles forces it to take stable paths only. The colliding particles quickly align at perfect $45^\circ$ lines, corresponding to the angles where the velocity field is radial only. The stability results from the steric interactions --- once the particles collide, the steric force cancels any velocity in the radial direction, and the particles slide on each other until the tangential velocity is also zero. Since the velocity field increases near the particle, the sliding motion is very fast compared to the global dynamics. Thus, a collision is characterized by a swift change to the Hamiltonian since the pair's contribution to $H$ changed from relatively constant (in time) to zero. We clearly see this rapid change in Fig.~\ref{fig:collision}, which shows a simulation for four particles. A  collision of two particles is clearly observed by the decrease in the Hamiltonian at $t=8.5$. In simulations of hundreds of particles, collisions are too frequent to see the effect clearly.

The two-particle behavior described above repeats on a global scale: all the particles collide into clusters, and all the clusters collide into bigger clusters. This is because, at large distances, a cluster functions as a particle whose strength is the combined strength of its constituents. Therefore, the dynamics of clusters can be described by a Hamiltonian in Eq. (\ref{eq:hemiltonian}), now with varying strengths. The center of activity is now given by $\sum_iS_i\textbf{r}_i/\sum_iS_i$ and is still conserved. In the case of two-cluster dynamics, the vector $\textbf{d} = \textbf{r}_1-\textbf{r}_2$ is sufficient to describe the dynamics, and its equation of motion is $\dot{\textbf{d}}=(S_1+S_2)\nabla^\perp\psi(\textbf{d})$. Therefore the two-cluster dynamics also lead to a collision, resulting in the clusters mimicking the behavior of the particles.
The collisions continue until all the particles form a single stable cluster in the form of an incomplete square lattice, see Fig.~\ref{fig:sim}. This final result can be deduced from the Hamiltonian, since, in each collision, the formed cluster must have zero Hamiltonian (i.e. its self interaction). 

The crystallization is thus a general behavior of any ensemble of identical particles with quenched orientation, as long as the two-particle dynamics lead to a collision, in the manner discussed in Sec.\ref{sec_twoGeneral}. Particles will tend to stabilize in incomplete crystal clusters, with an angle approximately given by the lowest power term of the stream function, which is dominant at close-range interactions. More precisely, the stability angles are such that the velocity field at the particles' radius ($l_{\text{steric}}/2$) at those angles is radial only. Of course, this generalized conclusion is true if the system does not find dynamic stability.

%As for interactions which do not lead to collisions, because their stream stream function has extrema, there is still aggregation, but particles will be separated by  larger distances given by the minimal points of the stream function. An example for such case will be the squirmer, introduced in Sec.\ref{sec_oneParticle}. Because the velocity field at boundary of the squirmer is only in the tangential direction, there are no stable paths that lead to a collision, but only paths like {\it a} (excluding paths with a zero Hamiltonian), which encircle a neutrally stable point (an extremum).

%We initialize 500 randomly positioned squirmers with the stream function $\psi \propto (\frac{R^2}{r^2}-\frac{R^4}{r^4})\sin(4\theta)$ with the same steric interactions mentioned earlier. No large structures were created, but instead the particles were spread out in a seemingly random distribution.
%Because the extremum points are only neutrally stable in the two-particle dynamics, they become unstable in the multi-particle dynamics, especially when they are many, hindering the possibility of dynamic global structure. In the SI, we briefly introduce an example with global dynamics, which arises from a stream function with less extrema.

In cases where the stream function is not a pure multipole and possesses extremum points, the dynamics do not necessarily result in the collision of two particles. The particles still draw closer up to distances where the next-order multipole term starts to dominant. Two examples of such cases are given in the SI. One still results in crystallization, with particles tending to stay at minimal points of the stream function. The second example is of squirmers. The particles draw closer, but within the aggregate, there are still chaotic dynamics. 

\section{Discussion}
In this work, we introduced a new method to characterize the flow created by 2D active particles, working directly with a stream function instead of the velocity field. We used this method to investigate the dynamics of a two-particle system, which we then used to infer the behavior of a multi-particle ensemble. We found that in all cases where the stream function does not have a local extremum and does not have a contribution from the monopole or the dipole terms, two particles will always collide. This includes all cases where the flow field is given by a pure even force moment (such as in the far-field limit). Our simulations suggest that many-body systems of such particles aggregate into an ordered cluster. The formed clusters appear to be fractal-like and are somewhat reminiscent of diffusion-limited aggregation \cite{witten1981diffusion} but with two differences --- (1) particle motion is driven by active flows, not diffusion. In fact, the only random component is in the initial positions of the particles. (2) There is short-range repulsion between particles, not attraction. Particles stick together due to the balance between steric interactions and active flows. 

The steps we took in this paper are not limited to a simple description of 2D fluid dynamics, but can be generalized to other effective 2D fluids with more complex flow equations. Such systems are commonly encountered in biological membranes \cite{saffman1975brownian, hosaka2017lateral, oppenheimer2019Crystalization, manikantan2020collective, bagaria2022dynamics}, or in experimental systems when particles sediment close to the bottom or float to the interface \cite{ben2022cooperation, lopez2014dynamics, tsang2014flagella, wioland2013confinement}. Such cases are associated with Green's functions that differ from Eq.~(\ref{eq:oseen}) and flow equations that differ from Eq.~(\ref{eq:main_DE}). Solutions to their flow equations will produce all the force multipoles of those systems. 

In addition, the arguments and results laid out in Sec.\ref{sec_twoGeneral}, regarding possible paths that two particles can take, are also relevant to these more complex systems, in describing particles whose orientation is frozen. The requirements for particle dynamics to result in a collision are:
(a) The stream function has no extrema, e.g. by the Hessian test (and twice-differentiability everywhere except the origin). 
%In particular, particle dynamics will lead to collision when: 
(b) $\psi(r \rightarrow \infty)=0$, and 
(c) there is at least one path diverging to infinity.

For instance, in a membrane, at small distances, the fluid is conserved in the membrane, and the flow created by an active protein is given by the dipole terms in Eq.~(\ref{eq:mega_solution}). At large distances, momentum is exchanged with the outer three-dimensional fluid \cite{saffman1975brownian, oppenheimer2009correlated}. An active particle in a membrane at large distances creates a flow with a stream function that dies out at infinity, as $\psi=S\frac{\sin2\theta}{r}$ \cite{manikantan2020collective}. The ray $\theta=0$ is a path diverging to infinity. The Hessian is negative, $\det\textbf{H}(\psi)=-2S(5+3\cos^2 2\theta)/r^6<0$. Thus, two oriented active proteins interacting via this stream function are bound to aggregate. 

We have pointed out a connection between two-particle dynamics and multi-particle dynamics. It is tempting to say that as long as $\psi \sim r^{-1}$ or lower, such that two particles must collide (per our discussion in Sec.\ref{sec_twoGeneral}), then larger systems will exhibit the same dynamics because the interactions are dominated by the nearest neighbors. However, it is unclear if this assessment is necessarily correct. Based on our findings, we hypothesize that such a conclusion is correct, but it remains for future work to assess its validity. 

%Indeed, we have discussed mainly the interactions leading to collisions in the two-particle dynamics. However, it is clear from our simulations, that complex dynamics are more likely to be found in other kinds of interaction --- those facilitated by a stream function that has minima and maxima. As hinted earlier, systems with many minima and maxima, will probably not result in a static final states. They are also less likely to have order in their dynamics. In fact, dynamic large structures and global properties are in all likelihood found in the middle of these extremes, such as the example in the SI. Such interactions require a closer examination.

We have focused on interactions that lead to collisions between two particles --- those of pure multipolar flows and cases where there are no extremum points in the stream function. 
In other types of dynamics, particles attract up to a certain distance, but there is no static final state. Ensembles with such interactions can have chaotic dynamics, but also global structures and order, as mentioned in the SI. These systems require closer examination.

We have limited our discussion to far-field hydrodynamic interactions in the low Reynolds limit. However, it is clear that in close range, the interaction of the particles can be influenced by other effects. For Instance, Yoshinaga and Liverpool \cite{lubrication_interaction} showed the importance of lubrication forces in dense interaction between active swimmers, which may affect the crystallization and stability of a many-particle ensembles. Similar effects are expected to change the final state of our system, but in principle, the characteristics of the far-range interactions should remain the same, meaning the particles are still expected to aggregate into large clusters.

\textbf{Acknowledgments}\\
We thank David B. Stein for his helpful comments. 
%\section*{Acknowledgments}

\bibliography{bib}% Produces the bibliography via BibTeX.

\vspace{0.5cm}
\section{Supplementary Material}

\subsection{Solving for the Streamfunction}
Here we develop the equation for the Fourier components of $\psi$ denoted by $B_n(r)$ from the flow equation given in the main text, and solve it. The right-hand-side is given by
%\lipsum[1]
%\begin{widetext}
\begin{align}
    \nabla^\perp\cdot\textbf{f}&=
    \frac{1}{R}\textbf{f}_0\cdot\nabla^\perp(\delta(r-R)F(\theta)) \\
    &=
    \frac{1}{R}\Re\left\{
    \eta e^{i\theta}\left[
    \frac{1}{r}\delta(r-R)\frac{\partial}{\partial\theta}
    -i\frac{\partial\delta}{\partial r}(r-R)
    \right]F(\theta)
    \right\},\nonumber
\end{align}
%\end{widetext}
%\lipsum[1]
where we used $\textbf{f}_0 = f_0(\hat{x}\cos\varphi + \hat{y}\sin\varphi) = \Re\left\{(\hat{r}+i\hat{\theta})\eta e^{i\theta}\right\}$. Plugging in the Fourier sums, the left-hand-side is identified with $-\hat{O}^2_nB_n$, as a result of the Laplacian in polar coordinates. Thus, the resulting equation is given by
\begin{equation}
\label{eq:semigreen}
    \hat{O}_n^2 B_n(r)=\frac{1}{R}i\eta C_{n-1}\left[\frac{d}{dr}\delta(r-R)-\frac{n-1}{r}\delta(r-R)\right],
\end{equation}
We ignore the $\Re$ operator in Eq. (1.1) because the solution is now given by the real part of $\psi$. To solve this equation, one must first solve the homogeneous equation $\hat{O}_n^2 B_n(r)=0$. This is an Euler equation whose solutions are always given by monoms. Here, after plugging in $B_n=r^m$ we get $m=\pm n, 2\pm n$. Thus, the solution to eq. (\ref{eq:semigreen}) is given by:

\begin{equation}
    B_n(r)=\frac{1}{R}\begin{cases}
x_{n}\left(\frac{r}{R}\right)^{|n|}+y_{n}\left(\frac{r}{R}\right)^{|n|+2} & r<R\\
z_{n}\left(\frac{r}{R}\right)^{-|n|}+w_{n}\left(\frac{r}{R}\right)^{-|n|+2} & r>R
\end{cases},
\label{eq:Bguess}
\end{equation}

as long as $n\neq0,\pm1$ which will be resolved later. The condition on the coefficients $x_n,y_n,z_n,w_n$ are given by the behavior at $r=R$. $B$ and $B'$ must be continuous at $r=R$, because their discontinuity would result in higher derivatives of $\delta(r-R)$ in the equation. To obtain discontinuity constraints on $B''$ and $B'''$ we first integrate over the equation from $R-\epsilon$ to $R+\epsilon$ and get one condition, and second, we multiply both sides by $(r-R)$ and integrate over the same interval, getting the second condition
\begin{equation}
    \left.rB''_n(r)\right|_{R-\epsilon}^{R+\epsilon}=i\eta C_{n-1},\left.r^2B'''_n(r)\right|_{R-\epsilon}^{R+\epsilon}=-i\eta C_{n-1}(n+1).
\end{equation}
After plugging in Eq.~(\ref{eq:Bguess}), we get equations for the coefficients
%\lipsum[1]
\begin{widetext}
\begin{equation}
    \left(\begin{array}{cccc}
1 & 1 & -1 & -1\\
|n| & |n|+2 & |n| & |n|-2\\
|n|(|n|-1) & (|n|+1)(|n|+2) & -|n|(|n|+1) & -(|n|-1)(|n|-2)\\
(|n|-1)(|n|-2) & (|n|+1)(|n|+2) & (|n|+1)(|n|+2) & (|n|-1)(|n|-2)
\end{array}\right)\left(\begin{array}{c}
x_{n}\\
y_{n}\\
z_{n}\\
w_{n}
\end{array}\right)=\left(\begin{array}{c}
0\\
0\\
-1\\
\frac{n+1}{|n|}
\end{array}\right)i\eta C_{n-1}R.
\end{equation}
\end{widetext}
%\lipsum[1]
The solution is 
\begin{align}
&z_{n}=\frac{i\eta C_{n-1}R}{4(1+|n|)}\left(\frac{1}{|n|}+\Theta(n)\right) \nonumber \\
&w_{n}=\frac{i\eta C_{n-1}R}{4(1-|n|)}\Theta(n),
\end{align}
where $\Theta$ is the step function. When $n=-1$ the solution can be gained from a limit in n, since $w_n=0$ for negative $n$. As mentioned in the main text, we don't concern ourselves with the case $n=1$, and so $n=0$ is the only special case. Its homogeneous solutions have a degeneracy, and so the general homogeneous solution is given as a linear combination of $1,r^2,\ln r,r^2\ln r$. Applying the same continuity conditions gives the solution in the main text, where a global constant is ignored.

\begin{figure*}[tbh]
    \centering
    \includegraphics[width=0.8\linewidth]{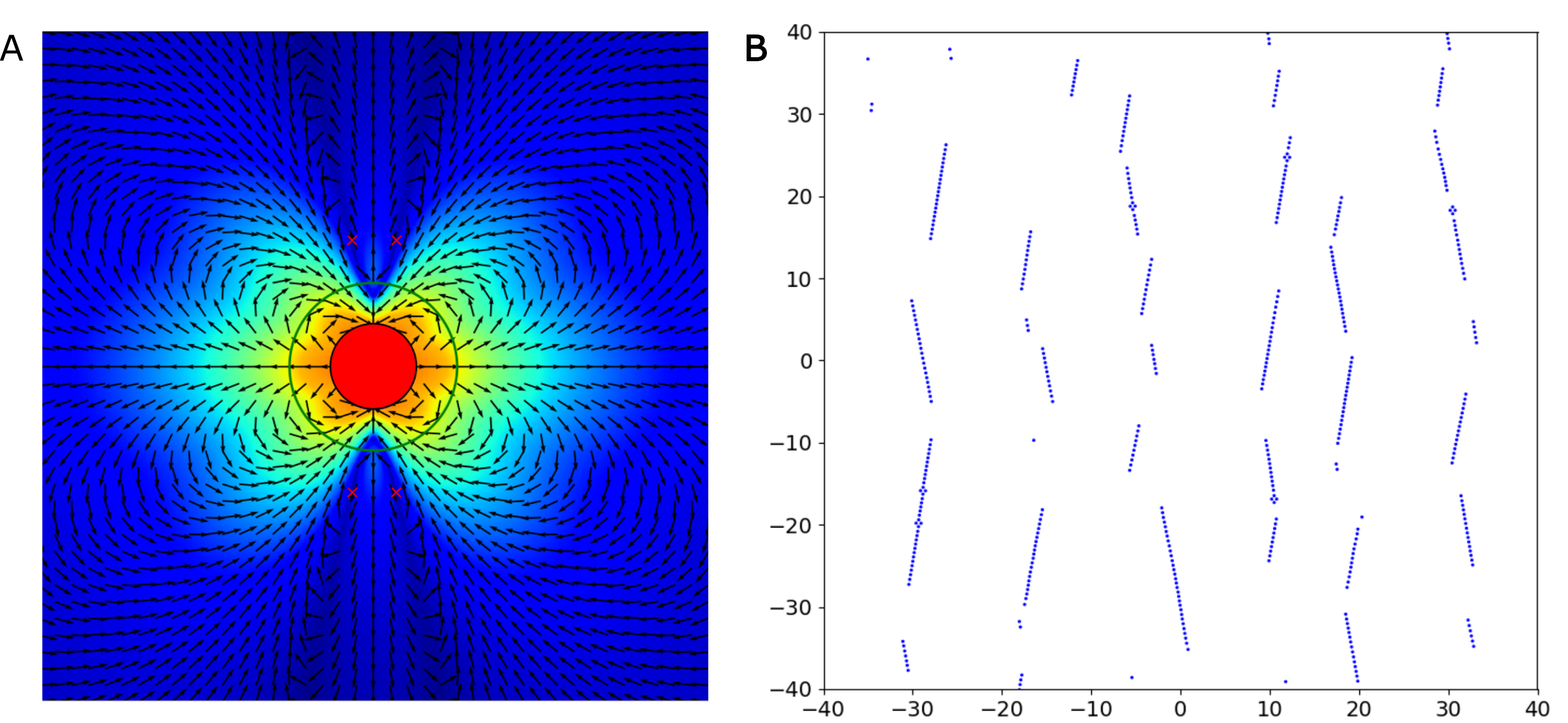}
    \caption{(a) The velocity field created by the stream function $\psi = (\frac{R^2}{3r^2}-\frac{R^4}{4r^4})\sin(4\theta)+\frac{R^2}{2r^2}\sin2\theta$. The red circle marks the particle. The green circle has twice its radius, noting the closest distance another particle can approach before steric interactions take effect. The red crosses mark the extremum points. (b) The final state of a simulation of particles with this stream function. The particles clearly form lines tilted at the angles of the extrema. In addition, a few particles collide and form diamonds, similar to the clusters presented in the main text. It is possible that at even longer times, all the particles will form a single line. }
    \label{fig:minimasThirdHarmonic}
\end{figure*}

\subsection{Constraints on the Force Distribution}
As mentioned in the main text, limiting the force distribution to a circle and to a constant direction does not constrain the solution of the flow given by the series solution. The coefficients in the series solution are given by:
\begin{equation}
    D_{ji_1i_2\dots i_n} = \frac{(-1)^n}{n!}\int_S f_jr_{i_1}\dots r_{i_n}dA
\end{equation}
where $S$ is the region the force is applied. The series solution is given by a sum of the contraction between $\textbf{D}_{n+1}$ and $\nabla^n\textbf{G}$, in the from $D_{ji_1\cdots i_n}\partial_{i_1}\cdots\partial_{i_n}G_{jk}$. Now, we will show that the possible leading terms in a solution are independent of $S$ in most cases, and indeed independent in our case, a circle. Because the permutation of $i_p,i_q$ doesn't change the component of either tensor, $\textbf{D}$ and $\nabla^n \textbf{G}$, the contraction is of the form:
\begin{equation}
\label{eq:contraction}
    \left(\begin{array}{c}
{n \choose 0}D_{i11\cdots11}\\
{n \choose 1}D_{i11\cdots12}\\
\vdots\\
{n \choose n}D{i22\cdots22}
\end{array}\right)\cdot\left(\begin{array}{c}
\partial_{1}^{n}\partial_{2}^{0}G_{ij}\\
\partial_{1}^{n-1}\partial_{2}^{1}G_{ij}\\
\vdots\\
\partial_{1}^{0}\partial_{2}^{n}G_{ij}
\end{array}\right)
\end{equation}
where in the first line all $i_p=1$, in the second line a single index is equal to $2$ and so on. We now denote $D_{i\alpha}$ the components of $\textbf{D}_n$ with $\alpha$ being the amount of $i_p=1$. Since the derivative of the green function does not depend on $S$, to show the result space of Eq. \ref{eq:contraction} does not depend on $S$, it is enough to show that the possible space of $D_{i\alpha}$ does not depend on $S$ (not a necessary but a sufficient condition). The components $D_{i\alpha}$ are defined as
\begin{equation}
    D_{i\alpha} = \frac{(-1)^n}{n!}\int_S f_i x^\alpha y^{n-\alpha} dA = \langle f_i,b_{\alpha}\rangle_S,
\label{eq:inner_product}
\end{equation}
where $\langle \cdot,\cdot\rangle_S$ is the standard inner product for function vector spaces, defined over $S$, and $b_{\alpha}=\frac{(-1)^n}{n!}x^\alpha y^{n-\alpha}$. First assume the direction of the force is fixed, say $f_2=0$. As long as $\mathcal{B}=\{b_{\alpha}|\alpha=1\cdots n\}$ are linearly independent under S, for every $\textbf{D}_n$ there exists $f_1\in\text{span}\mathcal{B}$ such that Eq. (\ref{eq:inner_product}) is satisfied. In particular, it is given by $f_1=\sum_{\alpha} D_{1\alpha}b^*_{\alpha}$, where $b^*_{\alpha}$ are the dual basis of $\mathcal{B}$. In particular, when $S$ is the unit circle, $b_{\alpha}=\cos^\alpha\theta\sin^{n-\alpha}\theta$ are linearly independent. Therefore, the possibilities of $\textbf{D}_n$ are not limited by the choice that $S$ is the unit circle, and therefore, the result of the series solution is not constrained by this choice. An example where the choice of $S$ does matter is when $\mathcal{B}$ is linearly dependent, for instance, if $S$ is a section of the line $y=x$.

However, the orientation of the force does matter. The solutions for the force that give a specific $D_{i\alpha}$ is $f_i=\sum D_{i\alpha}b^*_{\alpha}$. Because this force does not have a constant direction (in general), it is not possible to use the argument above here. 

However, a solution for a general force distribution is given by the sum of two familiar solutions --- one considering only the $x$-component of the force, and second only with the $y$-component. Therefore, the most general force distribution is given as a sum of the same terms discussed in the main text ($\psi_{nm}$ in Sec. IIIB), just with different coefficients. As a result, the discussion on the dynamics in the main text applies to any force distribution. An example is given in the following section.

\subsection{Constructing Multipoles with Point Forces}
Here we highlight how to construct force distributions using point forces only, such that a specific multipole can be dominant. As mentioned in the main text, the term $r^{-n}$ ($n>0$) couples to the harmonics $\pm n,n+2$ in the streamfunction. These harmonics couple to the harmonics $\pm n-1,n+1$ of the force distribution. Therefore if the lowest (positive) harmonic in a force distribution is $n+1$ (which is accompanied by $-n-1$), then $r^{-n}$ will be the leading term in the stream function.

\begin{equation}
\label{eq:n_term_psi}
    \psi = \frac{R^{n+1}}{4r^n} \Re\left\{
    -\frac{i\eta C_{n+1}}{n+1}e^{i(n+2)\theta}+
    \frac{i\eta C_{-n-1}}{n(n+1)}e^{-in\theta}
    \right\},
\end{equation}

where $C_{-n-1}=C_{n+1}^*$ because the force distribution is real. In order to accomplish Eq. (\ref{eq:n_term_psi}) with point forces, a force distribution is given by 
\begin{equation}
    F(\theta)=\sum\limits_{k=0}^{2n+1}(-1)^{k}\delta\left(\theta-\theta_{0}+\frac{\pi k}{n+1}\right).
\label{eq:force_dist}
\end{equation}
Intuitively, this function has a period of $2\pi/(n+1)$ and so its smallest harmonic must be $n+1$, and one can verify that using the Fourier expansion of the delta function, as well as the fact that the roots of $f(z)=1-z+z^{2}-\cdots-z^{2n+1}$ are all the $2n+2$ roots of unity except $-1$. This distribution is used in the main text in Sec. IIIA with $n=2$.

As mentioned in the main text, for $n=2$ this distribution creates both second and forth harmonics in $\psi$. As mentioned earlier, using a force distribution with a varying direction, it is possible to achieve more varied linear combination of the terms. For example, we can sum two force distributions, each in a constant direction, in order to isolate only one of the harmonics. The distribution in Eq. (\ref{eq:force_dist}) with $n=2$ gives (as noted in the main text with $\theta_0=0,\textbf{f}_0=f_0\hat{x},C_3=\frac{3}{\pi}$)
\begin{equation}
    \psi_\text{oct}=\frac{R^3}{4\pi r^2}f_0 \left(\sin4\theta +\frac{1}{2}\sin2\theta\right).
\label{eq:force_example}
\end{equation}
Now, to isolate only the forth harmonic, we rotate the force distribution (and $\textbf{f}_0$) by $\pi/2$, creating a distribution with $\textbf{f}_0||\hat{y}$. These two distributions can be summed to create a distribution with a varying force direction. Summing Eq. (\ref{eq:force_example}) with itself after the transformation $\theta\rightarrow\theta+\frac{\pi}{2}$ gives
\begin{equation}
    \psi=\frac{R^3}{2\pi r^2}f_0 \sin4\theta.
\end{equation}
Using this decomposition, any term we have mentioned $\psi_{nm}$ can be dominant. 

\subsection{Stream Functions with Extrema}
As mentioned in the main text, a simple example of a stream function with minima and maxima is given by a squirmer with $\psi = (\frac{R^2}{r^2}-\frac{R^4}{r^4})\sin(4\theta)$, which has extrema at $r=\sqrt{2}R$ and $\theta = \frac{\pi}{8},\frac{3\pi}{8}\cdots\frac{15\pi}{8}$. In close range, the particles seem to exhibit mostly chaotic dynamics, with weak stability around the stream function's extrema.

Another example is given by the stream function $\psi = (\frac{R^2}{3r^2}-\frac{R^4}{4r^4})\sin(4\theta)+\frac{R^2}{2r^2}\sin2\theta$, generated by a particle with only a third harmonic in its force distribution, but with a varying force direction. Its velocity field is displayed in Fig.\ref{fig:minimasThirdHarmonic}. In a two-particle interaction, it is clear that most paths that $\mathbf{d}$ takes lead to the $y$-axis, and very few paths (that begin very close the the particle) result in collision and static stability. Once $\mathbf{d}$ reaches the $y$-axis, small perturbations will change its paths into orbits around the extrema points. Thus, particles are likely to be stable around the extrema in multi-particle interactions.

Indeed, simulating a multi-particle interaction, in a similar manner to the main text, shows dynamical stability in the formation of rods, angled according to the angle of the extrema, at $\approx\pm9^\circ$ measured from the $y$-axis. These structures, while relatively stable, are still much easier to change than the structures created by attraction of two particles at their boundary. In fact, while the rods can scatter and destroy each other, the rare crystallizations in this simulation remain uninterrupted.

These examples show that when extrema are present, and crystallization is not forced, the dynamics can be very ordered or very chaotic, and a closer look is needed to understand close-range dynamics in general.

\end{document}